\documentstyle[12pt]{article}

\newcommand{\sect}[1]{\setcounter{equation}{0}\section{#1}\indent}

\begin{document}

\topmargin 0pt
\oddsidemargin 5mm
\def\bbox{{\,\lower0.9pt\vbox{\hrule \hbox{\vrule height 0.2 cm
\hskip 0.2 cm \vrule height 0.2 cm}\hrule}\,}}

\newcommand{\EQ}{\begin{equation}}
\newcommand{\EN}{\end{equation}}
\newcommand{\bea}{\begin{eqnarray}}
\newcommand{\ena}{\end{eqnarray}}
\newcommand{\hs}[1]{\hspace{#1 mm}}
\renewcommand{\a}{\alpha}
\renewcommand{\b}{\beta}
\renewcommand{\c}{\gamma}
\renewcommand{\d}{\delta}
\newcommand{\e}{\epsilon}
\newcommand{\shalf}{\frac{1}{2}}
\newcommand{\pa}{\partial}
\newcommand{\tri}{{\small $\triangle$}}
\newcommand{\dz}{\frac{dz}{2\pi i}}
\newcommand{\ra}{\rangle}
\newcommand{\lan}{\langle}
\newcommand{\nn}{\nonumber \\}
\def\a{\alpha}
\def\b{\beta}
\def\g{\gamma}
\def\G{\Gamma}
\def\d{\delta}
\def\D{\Delta}
\def\e{\epsilon}
\def\ve{\varepsilon}
\def\z{\zeta}
\def\t{\theta}
\def\vt{\vartheta}
\def\r{\rho}
\def\vr{\varrho}
\def\k{\kappa}
\def\l{\lambda}
\def\L{\Lambda}
\def\m{\mu}
\def\n{\nu}
\def\o{\omega}
\def\O{\Omega}
\def\s{\sigma}
\def\vs{\varsigma}
\def\S{\Sigma}
\def\vphi{\varphi}
\def\av#1{\langle#1\rangle}
\def\pa{\partial}
\def\na{\nabla}
\def\hg{\hat g}
\def\un{\underline}
\def\ov{\overline}
\def\tr{{\rm tr}}
\def\to{\rightarrow}
\def\sabs{\s_{\rm abs}}
\def\ceff{c_{\rm eff}}
\newcommand{\AP}[1]{Ann.\ Phys.\ {\bf #1}}
\newcommand{\NP}[1]{Nucl.\ Phys.\ {\bf #1}}
\newcommand{\PL}[1]{Phys.\ Lett.\ {\bf #1}}
\newcommand{\CMP}[1]{Comm.\ Math.\ Phys.\ {\bf #1}}
\newcommand{\PR}[1]{Phys.\ Rev.\ {\bf #1}}
\newcommand{\PRL}[1]{Phys.\ Rev.\ Lett.\ {\bf #1}}
\newcommand{\PTP}[1]{Prog.\ Theor.\ Phys.\ {\bf #1}}
\newcommand{\PTPS}[1]{Prog.\ Theor.\ Phys.\ Suppl.\ {\bf #1}}
\newcommand{\MPL}[1]{Mod.\ Phys.\ Lett.\ {\bf #1}}
\newcommand{\IJMP}[1]{Int.\ Jour.\ Mod.\ Phys.\ {\bf #1}}
\newcommand{\CQG}[1]{Class.\ Quant.\ Grav.\  {\bf #1}}
\newcommand{\PRep}[1]{Phys.\ Rep.\ {\bf #1}}
\newcommand{\RMP}[1]{Rev.\ Mod.\ Phys.{\bf #1}}

\begin{titlepage}
\setcounter{page}{0}

\begin{flushright}
COLO-HEP-379 \\
hep-th/9611189 \\
November 1996
\end{flushright}

\vspace{5 mm}
\begin{center}
{\large Radiation from a class of string theoretic black holes}

\vspace{10 mm}

{\large S. P. de Alwis\footnote{e-mail: dealwis@gopika.colorado.edu}
and K. Sato\footnote{e-mail: sato@haggis.colorado.edu}}\\
{\em Department of Physics, Box 390,
University of Colorado, Boulder, CO 80309}\\
\vspace{5 mm}
\end{center}
\vspace{10 mm}

\centerline{{\bf{Abstract}}}
The emission  of a scalar with low energy $\o$, from a
$D~(4\le D\le 8 )$ dimensional black hole with $n$ charges is
studied in both string and semiclassical calculations.
In the lowest order in $\o$, the weak coupling string and 
semiclassical calculations agree provided  that the
Bekenstein--Hawking formula is valid and the effective central charge
$\ceff=6$ for any $D$.
When the next order in $\o$ is considered  however,
there is no agreement between the two schemes unless $D=5$, $n=3$
or $D=4$, $n=4$.

\vspace{29 mm}
\end{titlepage}

\newpage
\renewcommand{\thefootnote}{\arabic{footnote}}
\setcounter{footnote}{0}

\setcounter{equation}{0}
\sect{Introduction}
\indent
It has been over twenty years since the Bekenstein--Hawking formula
of black hole entropy was presented, but until very recently there
has been no precise statistical mechanics explanation of this.
The possibility of a microscopic explanation of black hole entropy in
terms of the counting of string states  was first suggested a few
years ago in \cite{sus} and was  further developed in \cite{sen}.
In the past year  we have seen, for certain black holes within the  
context of type II superstring theory, that the number of string 
states correctly gives the black hole entropy geometrically 
calculated using the Bekenstein--Hawking formula.
First, this agreement was shown for a five dimensional ($D=5$) black 
hole carrying three types of charge ($n=3$) in the extremal 
configuration \cite{sv}.
Then it was extended  to non-extremal configurations
\cite{cm}--\cite{blmpsv}.
The agreement was also confirmed for $D=4$, $n=4$ black hole in both
extremal and non-extremal configurations \cite{ms0}--\cite{hlm}.
The analysis in string theory is based on D-brane technology at weak
coupling whereas black hole configurations are studied in the strong 
coupling regime.
In the case of extremal black holes, which correspond to BPS
saturated bound states of D-branes, there is a topological 
invariance which preserves the degeneracy of states as the coupling 
is varied.  
This is not the case, however, for non-BPS string configurations and non-extremal black holes.\footnote{Recently, Maldacena showed that at sufficiently low energies the weak coupling calculation agrees with strong coupling black hole calculation for a class of near extremal five-dimensional black holes \cite{mal2}.}

String theory knows not only about thermodynamic quantities such as
entropy, but also about dynamical processes such as emission from a 
black hole.
Emission processes inevitably include a non-extremal configuration in
which a black hole  emits, for example, scalar particles and falls
into an energetically stable extremal configuration.
Das and Mathur \cite{dm2} calculated the emission
rate of a scalar of energy $\o$ from a system of D-branes which
corresponds to the $D=5$, $n=3$ black hole and showed
that the string calculation and the semiclassical black hole
calculation agree under the condition, $T_R\ll T_L$.
($T_R$, $T_L$ are the temperatures of right
moving modes and left moving modes, respectively.)
Maldacena and Strominger \cite{ms} dropped the condition and showed  
in a semiclassical calculation that the absorption cross section 
factorizes as,
\EQ
\sabs\propto{\rho({\o/2T_L})\rho({\o/2T_R})\over\rho({\o/ T_H})},
\EN
where $\rho(x)=(e^x-1)^{-1}$ is the Bose distribution function,
and $T_H$ is the Hawking temperature.
They further showed that the corresponding emission rate,
\EQ
\G=\sabs\rho\left({\o\over T_H}\right){d^4k\over(2\pi)^4},
\EN
agrees with the string calculation.
In this agreement between the semiclassical calculation and the
string calculation, the factorizability of the absorption cross
section in the semiclassical calculation is essential.
(The factor $\rho(\o/2T_L)\rho(\o/2T_R)$, which is the product of the
distribution of left moving modes and that of the right moving ones,
naturally appears in the string calculation.)

In order to extend the discussion to other values of $D$ and $n$, let
us introduce the parameter
\EQ
\l={D-2\over D-3}-{n\over2}.
\EN
$\l$ becomes zero only when $D=5$, $n=3$ and $D=4$, $n=4$ if $4\leq
D\leq 8$.
These two combinations of spacetime dimension and the number of
charges are special because the corresponding black holes have
a regular event horizon and finite horizon area in the extremal
limit \cite{klopv}\cite{cy}.

In this paper we try to generalize the results explained above to the
$\l\neq0$ case to see whether an effective string model can describe 
the properties of black holes in arbitrary ($4\leq D\leq8$) 
dimensions with different numbers of charges. 
A priori one might expect this to work because the 
Bekenstein--Hawking entropy is $O(\hbar^{-1})$ and that is precisely
the behavior of the entropy of a one dimensional  gas of massless  
particles. 
As in \cite{hkrs} the emission rate at lowest order in the energy is  
compared with the semiclassical calculation of
Hawking radiation by using the result of  \cite{dgm}.
Now while Das and Mathur's work \cite{dm2} is based on D-brane  
technology, however, one can formulate the entire calculation of 
emission rate and absorption cross section for a $D$-dimensional 
black hole by considering an effective string model.
For the purposes of this paper we do not need to specify how this  
effective string is built up of configurations of D-branes, 
NS-NS-branes, M-branes etc.
We find that the Bekenstein--Hawking formula is recovered provided  
that the effective central charge $\ceff$ is six.\footnote
{The same conclusion has been reached by Halyo et al 
\cite{hkrs}\cite{hrs} for all black holes.}
Since $\ceff=6$ corresponds to the physical degrees of freedom in six
dimensional string theory whilst the above is true for general  
$D,~4\le D\le 8$, this result is somewhat puzzling.
In the $D=5$, $n=3$ black hole which is described in terms of 
D-1-branes bound to D-5-branes there is some understanding of this 
effective value of the central charge \cite{mal}. 
A similar argument has also been given in the case
of $D=4$, $n=4$ black hole \cite{mal} though there it appears to be  
more tentative.\footnote{For a different perspective, see \cite{tsey} 
and \cite{lw}.}
However in the general case it is  unclear why one should have this  
value for the effective central charge.

Since the string calculation of emission rate can be generalized to
$D$-dimensions, it is natural to ask whether one can also generalize
the   calculation of \cite{ms} to
$D$-dimensions, i.e., whether the absorption cross section
factorizes in arbitrary $D$-dimensions as it does in five
dimensions.\footnote{The absorption cross section factorizes also in
$D=4$, $n=4$ case \cite{gk}.}
We show that  the factorization does not occur unless $\l=0$.
This implies that the string calculation and the semiclassical
calculation do not agree when $\l\neq0$ beyond the lowest order in  
$\o$.
Perhaps this is an indication that the model makes sense only in the 
$\l =0$ case where indeed there is some explanation of why $\ceff=6$. 
However we hesitate to draw such a strong conclusion
since there is no reason why one should believe that the weak  
coupling string calculation should agree beyond extremality even in 
the $\l=0$ case.\footnote{See footnote 1.}  
It was a surprise even there and the agreement of the lowest order 
in $\o$ calculation for $\l\ne 0$ is also a surprise.  
What is not a surprise is the fact that it does not agree beyond 
the lowest order.

In the next section, we introduce our model.
We also derive some thermodynamic relations for later use.
In section 3, we calculate the emission rate and the absorption cross
section from a  string configuration and from a black hole in
$D$-dimensions, and see that the two results agree in the lowest
order in energy if $\ceff=6$ irrespective of $D$.
In section 4, higher order contributions are considered and it is
shown that the string calculation and the semiclassical calculation
agree only when $\l=0$.
We summarize our conclusions in the last section.

\sect{The Effective String Model}
\indent
We consider a string configuration in type II (either A or B) theory,
described by  the sum of the world
sheet $\s$-model action,
\EQ
I={1\over 2\pi\a'}\int d^2\s{1\over2}\sqrt{\g}\g^{\a\b}\pa_\a
X^\mu\pa_\b X^\nu
g_{\mu\nu},
\label{sigmamodel}
\EN
and the low energy effective action,
\EQ
S={1\over2\k^2}\int d^{10}x\sqrt{-g}e^{-2\phi}[R+\cdots].
\label{loweneff}
\EN
The $\s$-model action $I$ describes an effective string source
for the background field configuration. 
We expand the metric around a black string background 
$g_{\mu\nu}^0=\eta_{\mu\nu}+O(\k^2)$ as
\EQ
g_{\mu\nu}=g_{\mu\nu}^0
+\sqrt{2}\k h_{\mu\nu}.
\label{bgm}
\EN
$\k$ is the ten-dimensional gravitational coupling
constant, which is proportional to the string coupling constant $g$.
The second term describes the propagation of the graviton.
At strong coupling, $g^0_{\mu\nu}$ should
be the black string background in which gravitons propagate.
At weak coupling, on the other hand, one can neglect 
$O(\k^2)$ term and background
spacetime becomes  flat with a small fluctuation represented by
the second term on the r.h.s. of eq.(\ref{bgm}), 
which describes the emission from the string.

Let us look at the mass formula for a string with one dimension
compactified on a circle of radius $R$ (or the circumference $L=2\pi
R$).
\EQ
E^2=Q_L^2+8\pi T N_L=Q_R^2+8\pi T N_R,
\EN
where
\EQ
Q_L^2=\left(n_wLT-{2\pi n_p\over L}\right)^2,
\quad Q_R^2=\left(n_wLT+{2\pi n_p\over L}\right)^2.
\EN
$T=1/2\pi\a'$ is the fundamental string tension, $n_w$ is the winding
number, $n_p$ is the Kaluza--Klein mode. 
The matching condition $N_L-N_R = n_wn_p$ must be satisfied. 
Following \cite{dm1} and \cite{malsus}, we make the
long string approximation,
\EQ
{N_{L,R}\over L^2 T}\ll 1.
\label{longstrapprox}
\EN
The meaning of this will become clear soon.
We split the energy $E$ into that for the left movers and that for
the right movers under the long string approximation, 
eq.(\ref{longstrapprox}),
\EQ
E_L={1\over2}(E-E_0+P)={2\pi\over n_wL}N_L+{\it O}(L^{-2}),
\label{el}
\EN
\EQ
E_R={1\over2}(E-E_0-P)={2\pi\over n_wL}N_R+{\it O}(L^{-2}),
\label{er}
\EN
where $E_0=n_wLT \gg E_{L,R}$ is the ground state energy, 
$P=2\pi n_p/L = E_L - E_R$ 
is the total momentum flowing along the string. $E_L$ and $E_R$
describe the energy associated with the left and right moving 
oscillator excitations of the string. 
After rescaling $n_w L\to L$, we have
\EQ
E=E_0+E_L+E_R=LT+{2\pi\over L}(N_L+N_R)+{\it O}(L^{-2}).
\label{strenergy}
\EN
Since the black hole at strong coupling is expressed in terms of the
string sitting at the origin, its mass should be given by 
$M=E\simeq LT$ in the lowest order.
The fact $E_0\gg E_{L,R}$ under the long string approximation means
that the oscillator excitations are small compared with the ground state
energy. 
Thus we can expect that the excitations are sparsely distributed 
on the long string and that there is little interaction between them.

In the weak coupling limit, the oscillator excitations can be
considered as a one dimensional gas moving on the long string with 
independent left and right moving modes. 
We summarize its thermodynamics in the rest of this
section.
The emission of the graviton following the interaction of left and  
right movers is an effect of order $\k$. (See eq.(\ref{bgm}).) 
In the lowest order, the right moving and the left moving sectors 
don't interact each other, but each sector is in thermal equilibrium 
at a temperature either $T_L$ or $T_R$. 
Thus any thermodynamic quantity $O$ can be split into two parts, 
$O_L$ and $O_R$, for the left and right moving sectors, respectively. 
The free energy of a gas of one bosonic species is given by $-{\pi}LT_{L,R}^2/12$. 
We are studying a supersymmetric system with $f$ bosonic species and 
$f$ Majorana-fermionic ones.
The effective central charge is $\ceff=3f/2$.
Thus the total free energy is given by
\EQ
F_{L,R}=-{\pi\over12}LT_{L,R}^2\cdot{3\over2}f.
\EN
The entropy is
\EQ
S_{L,R}=-{\pa F_{L,R}\over \pa
T_{L,R}}={\pi\over6}LT_{L,R}\cdot{3\over2}f.
\EN
The energy of the system is
\EQ
E_{L,R}=F_{L,R}+T_{L,R}S_{L,R}={\pi\over12}LT_{L,R}^2\cdot{3\over2}f.
\EN
From the expressions above, we have the relations between $T_{L,R}$,
$E_{L,R}$ and $S_{L,R}$;
\EQ
T_{L,R}=\sqrt{8E_{L,R}\over \pi f L}={4S_{L,R}\over\pi f L}.
\label{tes}
\EN
When the total momentum is fixed, i.e., $\d P=\d E_L-\d E_R=0$, we
see $\d E\vert_P=\d E_L+\d E_R=2\d E_L=2\d E_R$. 
Therefore we have,
\EQ
{1\over T_H}={\pa S\over\pa E}
={\pa S_L\over2\pa E_L}+{\pa S_R\over2\pa E_R}
={1\over2}\left({1\over T_L}+{1\over T_R}\right).
\label{hlr}
\EN
There is one more useful relation, which is obtained from
eqs.(\ref{tes}) and (\ref{hlr}),
\EQ
T_LT_R={2\over\pi f L}T_H S.
\label{lrh}
\EN
Finally, from eqs.(\ref{el}), (\ref{er}) and (\ref{tes}), we recover
the well known expression,
\EQ
S=S_L+S_R=2\pi\sqrt{\ceff\over6}\left(\sqrt{N_L}+\sqrt{N_R}\right).
\EN

\eject
\sect{Lowest Order Analysis}
\subsection{String Calculation}
\indent
Following \cite{dm2}, we consider an emission process in which a  
right mover with momentum $p$ and a
left mover with momentum $q$ collide and emit a scalar into the
($D+1$)-dimensional space, coming from graviton polarized in the
compact directions, with momentum $k$. 
The $S$-matrix element for this process is
\EQ
S_{fi}=(2\pi)^2\d^2(p+q-k){-iA\over\sqrt{(2p_0L)(2q_0L)(2k_0V_9)}}.
\label{smatrix}
\EN
$V_9$ is the nine-dimensional spatial volume. 
$A$ is the amplitude determined by $I+S$.
We rescale $X^\mu$ to $X^\mu=\sqrt{2\pi\a'}{\bar X}^\mu$ in order to
get the standard normalization of the kinetic term. 
Also we set the world sheet to be flat, i.e., $\g^{\a\b}=\d^{\a\b}$,
\EQ
I=\int{1\over2}\pa_\a {\bar X}^\mu\pa_\a {\bar X}^\nu\eta_{\mu\nu}
+\k\int\pa_\a {\bar X}^\mu\pa_\a {\bar X}^\nu h_{\mu\nu}.
\label{intermed}
\EN
We consider the emission of graviton which is polarized, for example,
in 6 and 7 directions (when $D=4$). 
We also rescale $h_{67}$ to $h_{67}={\bar h}_{67}/\sqrt{2}$, 
which gives the correct normalization of the kinetic term
for ${\bar h}_{67}$ in the low energy effective action, $S$,
eq.(\ref{loweneff}). 
This modifies the second term in eq.(\ref{intermed}),
\EQ
\sqrt{2}\k\int\pa_\a {\bar X}^6\pa_\a {\bar X}^7{\bar h}_{67}.
\EN
This fixes the strength of the interaction and gives, to lowest  
order, the amplitude, $A=\sqrt{2}\k p\cdot q$, as in \cite{dm2}.
Thus we recover the $S$-matrix element given in \cite{dm2} when
comparing to IIB compactified to five dimensions.
Note that we have shown the duality between fundamental string and
D-string by explicitly showing that the two formulations give the 
same $S$-matrix element.
We consider a $D$-dimensional spacetime obtained by the
compactification on $T^{9-D}\times S_1$ with volume $V_{9-D} L$. 
The string of length $L$ is wrapped around $S_1$. 
One can calculate the emission rate of the scalar of energy $\o$.
The calculation is given in \cite{dm2} for $D=5$ and the
generalization to $D$-dimensions is straightforward.
One gets,
\EQ
\G_{\rm str}
={L\k_D^2\over4}\o\rho\left({\o\over2T_L}\right)
                  \rho\left({\o\over2T_R}\right)
                  {d^{D-1}k\over(2\pi)^{D-1}}.
\label{strerate}
\EN
Here, $\k_D^2=\k^2/LV_{9-D}=8\pi G_D$, $G_D$ being the 
$D$-dimensional Newton's constant.
The luminosity is given by
\bea
P_{\rm str}
&=&{L\k_D^2\over4}\o^2\rho\left({\o\over2T_L}\right)
                      \rho\left({\o\over2T_R}\right)
                      {d^{D-1}k\over(2\pi)^{D-1}}
\label{strluminosityomega}
\\
&\simeq&
{16G_D S\over f}T_H|K|^{-2}{d\o\over2\pi}\qquad(\o\to 0).
\label{strluminosity}
\ena
In the last step, we expanded the Bose distribution functions in
small $\o$ and used eq.(\ref{lrh}).
We also used the notation,
$|K|^2=(2\pi)^{D-2}\o^{-(D-2)}\O_{D-2}^{-1}$, where
$\O_{D-2}=2\pi^{(D-1)/2}/\G((D-1)/2)$ is the $(D-2)$-dimensional area
of the unit $(D-2)$-sphere.

Next, we consider an absorption process.
The five dimensional case was worked out in \cite{dm3}.
One can remove the restriction $T_R\ll T_L$ used in \cite{dm3} and
generalize the calculation to the $D$-dimensional case.
Since this is exactly the inverse process of the emission process
considered above, we use the same notations.
A scalar in ($D+1$)-dimensions with energy $\o$ and no momentum
in the string direction, polarized in ($9-D$)-dimensional
compact directions, is absorbed by a string, creating one left moving
excitation with momentum $p$ and one right moving excitation with
momentum $q$.
From the momentum conservation on the string world sheet,
$p_0=q_0=\o/2$.
The absorption cross section is defined by
\EQ
\sabs=2{\cal R}{\cal F}^{-1},
\EN
where ${\cal R}$ is the absorption rate, 
${\cal F}=\rho(\o/T_H)V_{D-1}^{-1}$ is
the flux of incident wave. 
$V_{D-1}$ is the volume of non-compact ($D-1$)-dimensional spatial 
directions transverse to the string. 
The factor 2 comes from the interchange of polarizations assigned 
on the two string excitations. 
Following \cite{dm3}, ${\cal R}$ is given by,
\EQ
{\cal R}={2\pi|R|^2\over \Delta E}
\rho\left({\o\over2T_L}\right)\rho\left({\o\over2T_R}\right).
\EN
We included the distribution functions of the right and left moving
excitations. $R$ is the amplitude to excite the string to any one of
the excited levels separated by $\Delta E=(2\pi/L)\times 2$ (see
eq.(\ref{strenergy})),
\EQ
R=\sqrt{2}\k|p_1|{1\over\sqrt{2\o L V_{9-D} V_{D-1}}}.
\EN
We split the nine-dimensional spatial volume $V_9$ into the product
of the length of the string, $L$, the volume of ($9-D$)-dimensional 
compact space, $V_{9-D}$, and the volume of ($D-1$)-dimensional 
tangential space, $V_{D-1}$.
Using $\k^2=L V_{9-D} \k_D^2$, one obtains
\EQ
\sabs={L\k_D^2\over4}\,\o\,
      {\rho(\o/2T_L)\rho(\o/2T_R)\over\rho(\o/T_H)}.
\label{sabsstr}
\EN

\subsection{Semiclassical Calculation}
\indent
Das et al \cite{dgm} studied a general $D$-dimensional metric of the
form
\EQ
ds_D^2=-f(r)dt^2+g(r)[dr^2+r^2d\O_{D-2}^2].
\EN
This metric can describe $D$-dimensional black holes with arbitrary
number of charges. 
They studied the propagation of a scalar field in this black hole
background. 
The absorption probability of s-wave is given by eq.(13) in \cite{dgm},
\EQ
|A|^2={A_H\over|K|^2},
\EN
from which one can calculate the luminosity,
\EQ
P_{\rm sc}={d\o\over2\pi}{\o\vert A\vert^2\over e^{\o/T_H}-1}
\simeq{d\o\over2\pi}T_H{\vert A\vert^2}
=A_HT_H|K|^{-2}{d\o\over2\pi}\qquad(\o\to 0).
\label{scluminosity}
\EN

By comparing it with the string calculation eq.(\ref{strluminosity}),
we obtain $A_H={16G_DS/ f}$. 
This implies $f=4$ (i.e. $\ceff=6$) irrespective of
$D$ in order for the Bekenstein--Hawking formula to hold.
This conclusion is very general.
No matter how many charges a black hole carries, and independent of
$D$ ($4\leq D\leq8$) the luminosity in the lowest order is reproduced 
by string theory calculation as long as $\ceff=6$.

\sect{Higher Order Contribution}
\indent
A natural question at this point is what happens if we take account
of not only the lowest order term in $\o$ but also the higher order 
terms?
For special cases, i.e., $D=5$, $n=3$ \cite{ms} and $D=4$, $n=4$
\cite{gk}, it was shown that the absorption cross section factorizes 
into Bose distribution functions,
\EQ
\sabs=|K|^2|A|^2
      ={A_HT_H\over4T_LT_R}\,\o\,
       {\rho(\o/2T_L)\rho(\o/2T_R)\over\rho(\o/T_H)},
      \qquad D=4, 5.
\label{sabssc}
\EN
Note that the factor $|K|^2$ converts the absorption probability
$|A|^2$ into the absorption cross section $\sabs$. 
(The former is calculated for a spherical wave and the latter is for 
a plane wave.)
So the emission rate is
\EQ
\G_{\rm sc}=\sabs\rho\left({\o\over
T_H}\right){d^{D-1}k\over(2\pi)^{D-1}}
={A_HT_H\over4T_LT_R}\,\o{\rho\left({\o\over2T_L}\right)
\rho\left({\o\over2T_R}\right)}{d^{D-1}k\over(2\pi)^{D-1}},\qquad
D=4, 5.
\label{msemissionr}
\EN
Eqs.(\ref{msemissionr}) and (\ref{strerate}) agree including the
constant factor, and so do eqs.(\ref{sabssc}) and (\ref{sabsstr}), 
because $L\k_D/4=A_HT_H/4T_LT_R$ by eq.(\ref{lrh}) and $A_H=16G_DS/f$.

This is the motivation for us to try to generalize \cite{ms} to
arbitrary $D$ and $n$.
In the semiclassical calculation in $D$-dimensions, we will calculate
the absorption cross section, which appears in eq.(\ref{msemissionr}).
In order to see an agreement with (\ref{sabsstr}), $\sabs$ must take
the following form
\bea
\sabs&=&{A_HT_H\over 4T_RT_L}\o
{\rho(\o/2T_L)\rho(\o/2T_R)\over\rho(\o/T_H)}
\label{compfactor}
\\
&=&A_H\left[1+{\o^2\over48}{1\over T_LT_R}+{\it O}(\o^3)\right],\quad
\o\to 0.
\label{factorizability}
\ena
The complete factorization as in (\ref{compfactor}) might be difficult 
to see because one has to include all the $\o$ dependence. 
Instead of eq.(\ref{compfactor}), we study the factorizability of 
$\sabs$ up to ${\it O}(\o^2)$ for small $\o$ as in 
eq.(\ref{factorizability}).

We study the propagation of a (neutral) scalar field $\phi$ in the
background of a $D$-dimensional black hole carrying $n$ charges,
\EQ
ds^2_D=-f^{-{D-3\over D-2}}hdt^2+f^{1\over D-2}(h^{-1}dr^2+r^2
d\O^2_{D-2}),
\label{metric}
\EN
where
\EQ
f=\prod_{i=1}^n\left(1+{r_i^{D-3}\over r^{D-3}}\right),
\qquad h=1-{r_0^{D-3}\over r^{D-3}}.
\EN
$r_i$ are charges. $r_0$ is the location of the horizon and it also
works as a non-extremality parameter.
($r_0=0$ corresponds to the extremal limit.)
This form of metric was first introduced by Cveti{\v c} and Tseytlin
\cite{ct}. (See also \cite{kt}.)
It covers various black hole configurations. 
For example, $n=0$ gives the Schwarzschild black hole in 
$D$-dimensions. 
$D=5$, $n=3$ and $D=4$, $n=4$ black holes are, of course, special 
cases of eq.(\ref{metric}) \cite{hms}.
As in the case of $D=5$, $n=3$ and $D=4$, $n=4$ black holes, this
$D$-dimensional black hole is assumed to be obtained from a
($D+1$)-dimensional black string by compactifying the string 
direction on a circle.
It is sometimes convenient to use another set of parameters, $\a_i$
and $\s$, instead of $r_1,\ldots,r_n$,
\bea
r^{D-3}_i&=&r^{D-3}_0 \sinh^2\a_i\quad (i=1,\ldots, n-1),
\\
r^{D-3}_n&=&r^{D-3}_0 \sinh^2\s.
\ena
In the black string picture, $r_n$ corresponds to the momentum
flowing along the black string, which is obtained by applying a 
boost to the string. 
The boost is parameterized by $\s$. 
In the extremal limit $r_0\to0$, $r_n$ is kept fixed, i.e., 
$\s\to\infty$.
We want to solve the wave equation for $\phi$ in the dilute gas
approximation \cite{ms},
\EQ
r_0,r_n\ll r_1,\ldots,r_{n-1}.
\label{dga}
\EN
Under this approximation, the metric (\ref{metric}) describes
a black hole with mass
\EQ
M=M_0+(D-3){\O_{D-2}\over2\k_D^2}{r_0^{D-3}\over2}\cosh2\s.
\EN
The first term
$M_0=[(D-1)r_0^{D-3}/2+(D-3)\{r_1^{D-3}+\cdots+r_{n-1}^{D-3}\}]
\O_{D-2}/2\k_D^2$ corresponds to the ground state energy $E_0$ in
eq.(\ref{strenergy}).
The second term corresponds to the oscillator excitation, $E_L+E_R$  
in eq.(\ref{strenergy}).
Using the $(D-2)$-dimensional area of the black hole horizon,
\EQ
A_H=A_{D-2}={D-3\over 4\pi}{r_0^{D-3}\over T_H}\O_{D-2},
\label{horizonarea}
\EN
one can calculate the Bekenstein--Hawking entropy,
\EQ
S=(D-3){\O_{D-2}\over2\k_D^2}{r_0^{D-3}\over T_H},
\EN
where the Hawking temperature $T_H$ is given by
\EQ
T_H={D-3\over 4\pi}r_0^{-(D-3)\l}\left({r_0\over r_1\cdots
r_{n-1}}\right)^{(D-3)/2}
{1\over\cosh\s}.
\label{hawktemp}
\EN
 From the black string point of view, the right moving oscillator
excitations and the left moving ones are separately in equilibrium at
temperatures;
\EQ
T_R={D-3\over 4\pi}r_0^{-(D-3)\l}\left({r_0\over r_1\cdots
r_{n-1}}\right)^{(D-3)/2}
e^\s,
\label{righttemp}
\EN
\EQ
T_L={D-3\over 4\pi}r_0^{-(D-3)\l}\left({r_0\over r_1\cdots
r_{n-1}}\right)^{(D-3)/2}
e^{-\s},
\label{lefttemp}
\EN
respectively.
They give the connection between string theory quantities on the
l.h.s. and black hole quantities on the r.h.s.
The relation, eq.(\ref{hlr}), is satisfied.
Note that one can determine the effective length of the string $L$  
from
eq.(\ref{lrh}),
\EQ
L={4\over\k_D^2}{4\over f}
  {\pi\over D-3}\left(r_0^{2\l}r_1\cdots
r_{n-1}\right)^{D-3}\O_{D-2}.
\EN

We look for a spherically symmetric configuration of a scalar field
$\phi(t,r)=e^{-i\o t}R(r)$, which satisfies the wave equation,
$\sqrt{-g}^{-1}\pa_\mu\sqrt{-g}g^{\mu\nu}\pa_\nu\phi=0$. $g_{\mu\nu}$
is given in eq.(\ref{metric}).
The radial equation becomes,
\EQ
\left[{h\over r^{D-2}} {d\over dr}hr^{D-2}{d\over dr}+\o^2
f\right]R(r)=0.
\label{radeq}
\EN
It seems extremely difficult, however, to solve it for arbitrary $r$.
Following \cite{ms}, we solve it separately in two regions, i.e.,
in the near zone, $r\leq r_m$, and in the far zone, $r\geq r_m$,
where the matching point $r_m$ is chosen to satisfy
\EQ
r_0,r_n \ll r_m \ll r_1,\ldots,r_{n-1}.
\EN
Also we impose the low energy condition \cite{ms},
\EQ
\o\ll{1\over r_1},\ldots,{1\over r_{n-1}}.
\label{lec}
\EN

In the far zone, eq.(\ref{radeq}) becomes,
\EQ
{d^2\psi\over d\rho^2}+\left[1-{(D-2)(D-4)\over4\rho^2}\right]\psi=0,
\EN
where $\rho=\o r$ and $R=r^{-(D-2)/2}\psi$.
It has two independent solutions
\EQ
F=\sqrt{\pi\over2}\rho^{1/2}J_{(D-3)/2}(\rho),
\EN
\EQ
G=\sqrt{\pi\over2}\rho^{1/2}N_{(D-3)/2}(\rho),
\EN
which are the Bessel and Neumann functions respectively.
The most general form of the solution is a linear combination of $F$
and $G$ with arbitrary constants $\a$ and $\b$,
\EQ
R=r^{-(D-2)/2}(\a F+\b G)
 =\sqrt{\pi\over2}\o^{1/2}r^{-(D-3)/2}
  \left[\a J_{(D-3)/2}(\o r)+\b N_{(D-3)/2}(\o r)\right].
\EN
In the limit $r\to\infty$, $R$ becomes
\EQ
R\simeq 2^{-1}r^{-(D-2)/2}
 \left[e^{-i\o r}\left(\a e^{ i(D-2)\pi/4}+\b e^{ iD\pi/4}\right)
      +e^{ i\o r}\left(\a e^{-i(D-2)\pi/4}+\b e^{-iD\pi/4}\right)
 \right].
\label{rinfty}
\EN
The first term in the square brackets is the incoming part while the
second term is the outgoing part. Near the matching point $r_m$,
\EQ
R\simeq\sqrt{\pi\over2}
 \left[{\a\o^{D-2\over2}
        \over2^{D-3\over2}\G\left({D-1\over2}\right)}
      +{\b\o^{1/2}\over\pi r^{D-3\over2}}
        \left\{{\o^{D-3\over2}\over2^{D-5\over2}}
               {\g+\ln(\o r/2)\over\G\left({D-1\over2}\right)}
               -{2^{D-3\over2}\G
               \left({D-3\over2}\right)\over(\o r)^{D-3\over2}}
        \right\}
 \right],
\EN
for $D$ odd, and
\EQ
R\simeq\sqrt{\pi\over2}
 \left[{\a\o^{D-2\over2}
        \over2^{D-3\over2}\G\left({D-1\over2}\right)}
       +\b{(-1)^{D+2\over2}\o^{1\over2}
        \over\G\left(D-1\over2\right)r^{D-3\over2}}
        \left({2\over\o r}\right)^{D-3\over2}
 \right],
\EN
for $D$ even.
The term multiplying $\a$ is common.
Near the matching point, the term multiplying $\b$ becomes 
divergent for both $D$ odd and even, under the low energy 
condition, eq.(\ref{lec}).

In the near zone,
\bea
f&\simeq&\left[\prod^{n-1}_{i=1}\left({r_i\over
r}\right)^{D-3}\right]
\left(1+{r_n^{D-3}\over r^{D-3}}\right)
\nn
&=& v^{n-1}\left({r_1\cdots r_{n-1}\over r_0^{n-1}}\right)^{D-3}
   +v^n    \left({r_1\cdots r_{n}  \over r_0^{n}  }\right)^{D-3}.
\ena
$v=(r_0/r)^{D-3}$ is a new radial variable. Eq.(\ref{radeq}) becomes
\EQ
(1-v){d\over dv}(1-v){dR\over dv}
 +v^{-2\l}\left(B+{C\over v}\right)R=0,
\label{diffeq}
\EN
where
\bea
B&=&C\left({r_n\over r_0}\right)^{D-3}=C\sinh^2\s,
\\
C&=&\left({\o r_0\over D-3}\right)^2\left({r_1\cdots r_{n-1}\over
r_0^{n-1}}\right)^{D-3}.
\ena
$B$ and $C$ are related to the temperatures, eqs.(\ref{hawktemp}),
(\ref{righttemp}) and (\ref{lefttemp}) by
\EQ
\sqrt{B+C}={1\over 4\pi}{\o\over T_H},\qquad
C=\left({\o\over 4\pi}\right)^2{1\over T_L T_R}.
\label{bcc}
\EN
Near the horizon, $v\simeq1$, we have the ingoing solution
\EQ
R_{\rm in}=e^{-i\sqrt{B+C}\log(1-v)}=(1-v)^{-i\sqrt{B+C}},
\label{ingosol}
\EN
which is independent of $\l$.

When $r$ is in the near zone but not very close to the horizon $r_0$,
we assume the form of the solution to be,
\EQ
R=R_0z^A F(z),\quad z=1-v.
\label{rzf}
\EN
We require that $F\to 1$ as $z\to 0$. Thus $A=-i\sqrt{B+C}$. (See
eq.(\ref{ingosol}).)
$R_0$ is a normalization constant to be determined by the matching
condition.
We look for a solution in series expansion
\EQ
F(z)=\sum^\infty_{n=0}b_nz^n,\qquad b_0=1.
\label{seriesexp}
\EN
Substituting eqs.(\ref{rzf}) and (\ref{seriesexp}) into
eq.(\ref{diffeq}), we
have a recurrence relation for $b_n$;
\EQ
n(n+2A)b_n=-\sum_{m=0}^{n-1}{1\over(n-m)!}
 \left[(k-1)_{n-m}B+(k)_{n-m}C\right] b_m \qquad (n\geq 1),
\label{recrel}
\EN
where $k=2\l+1$.
$(k)_n$ is the Pochhammer symbol; $(k)_n=k(k+1)\cdots(k+n-1)$,
$(k)_0=1$. 
We expand $b_n$ in powers of $\o$. $b_n^{(m)}$ denotes a term of 
order $O(\o^m)$. 
Note $A\sim{O}(\o)$ and $B\sim C\sim{O}(\o^2)$.
Up to second order in $\o$, eq.(\ref{recrel}) becomes
\EQ
b_n^{(0)}+b_n^{(1)}+b_n^{(2)}=
-\sum_{m=0}^{n-1}{1\over
n^2(n-m)!}\left[(k-1)_{n-m}B+(k)_{n-m}C\right]
b_m^{(0)} \qquad (n\geq 1).
\EN
By comparing both sides of the equation order by order, we get
$b_n^{(0)}=0,
b_n^{(1)}=0, b_n^{(2)}=-B(k-1)_n/n^2n!-C(k)_n/n^2n!\quad(n\geq1)$,
or
\bea
b_0&=&1,
\\
b_n&=&-{(k-1)_n\over n^2n!}B-{(k)_n\over n^2n!}C
    +{O}(\o^3)\qquad(n\geq1).
\ena
$R$ behaves near the matching point, $z\simeq1$ (i.e. $v\simeq0$), as
\EQ
R=R_0\sum^\infty_{n=0}b_n+{O}(v),\quad v\to 0.
\EN
The sum of $b_n$ takes the following form;
\EQ
\sum^\infty_{n=0}b_n
=1-B(k-1)\sum_{n=1}^{\infty}{(k)_{n-1}\over n^2n!}
        -C\sum_{n=1}^{\infty}{(k)_n\over n^2n!}    +{O}(\o^3).
\label{rdc}
\EN
The matching condition is given by
\EQ
R_0\sum_{n=0}^\infty
b_n=\sqrt{\pi}\a\left({\o\over2}\right)^{D-2\over2}
\left[\G\left({D-1\over2}\right)\right]^{-1}.
\label{mc}
\EN
We dropped the $\b$-dependent term by requiring $\a\gg\b$. 
In order to calculate the absorption probability, we study the 
conserved flux, which is defined by
\EQ
f={1\over2i}\left[R^*hr^{D-2}{dR\over dr}-{\rm c.c.}\right].
\EN
The incoming flux from infinity is given by the first term of
eq.(\ref{rinfty}),
\EQ
f_{\rm in}=-{\o\over4}|\a|^2.
\label{fin}
\EN
The absorbed flux is calculated from eqs.(\ref{rzf}) and
(\ref{seriesexp}) at the horizon, $z\simeq1$. 
One gets
\EQ
f_{\rm abs}=-{D-3\over4\pi}{r_0^{D-3}\over T_H}\o|R_0|^2.
\label{fabs}
\EN
We used eq.(\ref{bcc}).
Absorption probability is computed from eqs.(\ref{mc}), (\ref{fin})
and
(\ref{fabs}),
\EQ
|A|^2={f_{\rm abs}\over f_{\rm in}}
     ={D-3\over4\pi}{r_0^{D-3}\over T_H}
      {\pi\o^{D-2}\over2^{D-4}}
      \left[\G\left({D-1\over2}\right)\right]^{-2}
      \left|\sum_{n=0}^\infty b_n\right|^{-2}.
\EN
Using the factor $|K|^2$, one can calculate the absorption cross
section,
\EQ
\sabs=|K|^2|A|^2=A_H\left|\sum_{n=0}^\infty b_n\right|^{-2}.
\label{sabsexp}
\EN
$A_H$ is given in eq.(\ref{horizonarea}).
Combining eq.(\ref{rdc}) with eq.(\ref{sabsexp}), one gets
\EQ
\sabs
    =A_H\left[1+2B(k-1)\sum_{n=1}^{\infty}{(k)_{n-1}\over n^2n!}
               +2C\sum_{n=1}^{\infty}{(k)_n\over n^2n!}
               +{\it O}(\o^3)\right].
\label{sabs1}
\EN
This should be compared with eq.(\ref{factorizability}),
\EQ
\sabs
    =A_H\left[1+2C\zeta(2)+{\it O}(\o^3)\right].
\label{sabs2}
\EN
We used eq.(\ref{bcc}) and $\zeta(2)=\pi^2/6$.

Notice that there is no $B$ (or $r_n$) dependence in
eq.(\ref{sabs2}).
Therefore, the $B$ term in eq.(\ref{sabs1}) must vanish;
\EQ
(k-1)\sum_{n=1}^{\infty}{(k)_{n-1}\over n^2n!}=0.
\label{nobdep}
\EN
This equation has at least one solution, $k=1$.

\noindent
0) When $k=1$, eq.(\ref{sabs1}) becomes $\sabs=A_H[1+2C\zeta(2)
+{\it O}(\o^3)]$, i.e., eq.(\ref{sabs2}) is satisfied.

\noindent
This corresponds to $\l=0$, i.e., $D=5$, $n=3$ and $D=4$, $n=4$ cases
\cite{ms}\cite{gk}.

If there is another value of $k_0$ which realizes the factorization,
it has to satisfy the following two equations,
\EQ
\sum_{n=1}^{\infty}{(k_0)_{n-1}\over n^2n!}=0,
\label{cond1}
\EN
\EQ
\sum_{n=1}^{\infty}{(k_0)_{n}\over n^2n!}=\zeta(2).
\label{cond2}
\EN
The first condition eliminates the $B$-dependent term in
eq.(\ref{sabs1}). 
The second condition gives the right coefficient to the ${O}(\o^2)$
term.

\noindent
1) When $k_0\geq0$ ($\l\geq-1/2$), eq.(\ref{cond1}) is not satisfied
because the l.h.s. becomes a sum of positive numbers.

\noindent
2) When $-1<k_0<0$ ($-1<\l<-1/2$), eq.(\ref{cond2}) is not satisfied
because $k_0<0$ and $k_0+1>0$ mean,
\EQ
\sum_{n=1}^{\infty}{(k_0)_{n}\over n^2n!}
=k_0\sum_{n=1}^{\infty}{(k_0+1)_{n-1}\over n^2n!}<0.
\EN

\noindent
3) When $k_0$ is a negative integer ($\l=-1, -3/2, -2,
-5/2,\ldots$), the sum in eq.(\ref{cond2}) becomes a finite sum of 
rational numbers, which cannot be equal to an irrational number 
$\zeta(2)$.

1) and 3) eliminate all the possibilities for $D=4$, 5 except for
$\l=0$ (from
result 0)). In fact,
\EQ
\l_{D=4}=2-{n\over2}=2, {3\over2}, 1, {1\over2}, 0, -{1\over2}, -1,\ldots.
\EN
\EQ
\l_{D=5}={3\over2}-{n\over2}={3\over2}, 1, {1\over2}, 0, -{1\over2}, -1,\ldots.
\EN

As far as we know, $k=1$ ($\l=0$) is the only possibility when
factorization occurs. 
For negative non-integral $k_0$'s, possibilities of factorization 
have not been eliminated. 
However, it seems unlikely that eqs.(\ref{cond1}) and 
(\ref{cond2}) are both satisfied.

\sect{Summary}
\indent
We have compared the string calculation at weak coupling with the
semiclassical calculation at strong coupling. 
The following  is a summary of what we have learned.

\noindent
a) From  the lowest order in $\o$ calculation, we found that the  
Bekenstein--Hawking entropy is reproduced for arbitrary $D$ 
(and $n$) provided that $\ceff=6$.

\noindent
b) When $\l=0$ (i.e., $D=5$, $n=3$ or $D=4$, $n=4$), there is  
agreement between the string calculation and the semiclassical 
one to all orders in $\o$ with $\ceff=6$ \cite{ms}\cite{gk}.

\noindent
c) For $\l>0$, $-1<\l<0$, $\l=-m/2\,(m=2, 3, 4,\ldots)$, the emission
rate in the string calculation and that in the semiclassical 
calculation don't agree beyond the lowest order in $\o$.

\vskip 0.5cm
\noindent
{\bf Acknowledgements:}\\
We thank S. Chaudhuri for stimulating conversations. This work was
partially supported by Department of Energy contract No.
DE-FG02-91-ER-40672.

\end{document}